# Photostimulated Aggregation of Metal Aerosols

Sergei V. Karpov, Ivan L. Isaev
*L. V. Kirenski Institute of Physics, Russian Academy of Sciences,
Krasnoyarsk 660036, Russia*

**Abstract**. The effect of optical radiation on the rate of aggregation of nanoscopic particles is studied in metal aerosols. It has been shown that under light exposure, polydisperse metal aerosols can aggregate up to two orders faster due to the size dependent photoelectron effect from nanoparticles. Different size nanoparticles undergo mutual heteropolar charging when exchanging photoelectrons through the interparticle medium to result in an increased rate of aggregation. It is shown that long-range electrostatic attractive forces drive the particles into closer distances where the short-range Van-der-Waals forces become dominating. Attention is drawn to the fact that this effect may occur in various types of dispersed systems as well as in natural heteroaerosols.

## 1. Introduction

Studying the ability of aerosols containing metallic or dielectric dispersed phases as well as heteroaerosols of natural disperse media to aggregate under light exposure is of great practical interest [1]. Photostimulated coagulation of disperse systems widely occurs in nature and affects other processes. In particular, aggregation of superdispersed particles of prevailing atmospheric aerosols can cause significant changes in their optical extinction spectra [2]. A correct computation of this effect enables one to assess the influence of absorption of solar radiation by dust particles in the air on the thermal balance in the atmosphere. Finding a solution to this problem may be related to problems of global climatology.

Understanding the mechanisms of photoaggregation is useful for fabrication and storage of nanomaterials with aggregative instability. It is also useful when dealing with different technological processes that can be accompanied by formation of polydisperse metal nanoparticles and by their undesirable photoaggregation, which happens for example in coating chambers, or in vacuum atomizer cells with metal vapors used for nonlinear optical frequency conversion of pulsed laser radiation.

Analysis of the influence of optical radiation on the rate of aggregation of heterogeneous aerosols in combination with the accompanying factors, such as external corpuscular fluxes under close to real conditions [3] (dusty planet atmospheres, interstellar medium, industrial air pollutants), can reveal general mechanisms underlying the aggregation kinetics and evolution of a disperse system.

It has to be admitted that the experimental facts on the effect of light on coagulation of aerosols are not given due consideration and they are usually ignored because of the lack of theoretical models for their explanation.

In our paper we study the role of optical radiation in coagulation kinetics of metal aerosols and seek to answer the question: How can light accelerate the aggregation of this type of disperse systems? We propose one of the possible approches to explanation of photostimulated aggregation of metal aerosols. Similarly to photoaggregation in metal hydrosols [4,5], our model is based on the photoelectric effect. Under certain conditions, the photoelectric effect can stimulate mutual heteropolar charging of aerosol particles, provided the particles are polydisperse [6]. Under these conditions, long-range electrostatic attractive forces between particles initiate their approach into the range of action of the short-range Van-der-Waals forces and cause the particles to coagulate. In addition, electrostatic interactions of the particles randomize their motion, which stimulates collisions.

Optical radiation can induce electrostatic interaction of electroconductive particles, subject to the following.

Firstly, the metal sol must be polydisperse. In real nanoparticle sols, there are always particles of various sizes. Characteristics of a particle, such as the Fermi energy, photoelectric work function [6-10], and quantum yield [8], are controlled by the size of the particle. Under electromagnetic irradiation, smaller particles of such an ensemble will emit more photoelectrons from unit surface area than larger particles due to higher photoelectron emission constants of smaller particles.

Secondly, the interparticle medium must be transparent for photoelectrons, i.e. this medium should not impede transfer of photoelectrons from one particle to another. Due to the exchange of photoelectrons, such a system comes to equilibrium once the particles of a smaller size become charged positively while the particles of a larger size acquire the negative charge (it was studied by Nagaev in [6] and in his earlier papers cited therein. This is the basic idea of the mutual heteropolar charging effect. In the process of electron exchange the system tends to equalize Fermi energies in different particles. If the characteristics of particles were independent of their sizes, the total capture of photoelectrons and their emission would be controlled only by the surface of particles, and mutual charging would be impossible.

Thirdly, the interparticle medium (electrically neutral on the whole) must have a fairly low concentration of electric charges so as not to allow them to condense into electric double layers on the surface of the particle thereby compensating the metal kernel's own charge. The most favorable case would be when there are no charges in the interparticle medium other than the charges of photoemissive origin.

The aim of the present paper is to study the conditions for accelerated aggregation of aerosols with electroconductive nanoparticles under light exposure and to fill the gaps in the theoretical concepts as well as to prove a feasibility of the effect of mutual heteropolar charging in polydisperse ensembles of metal nanoparticles and to stimulate experimental studies of these effects in aerosols.

The paper is organized as follows. In Section 2, we describe the theoretical model, which includes the conditions for size-dependent photoelectron emission from nanoparticles, capture of photoelectrons and charging of particles, and also the basic equations of the molecular dynamics method used in studying the kinetics of aggregation under interparticle electrostatic and van-der-Waals interactions. In Section 3, we present the results of computation of the coagulation kinetics employing the model described in the previous Section and analyze the influence of different characteristics of the disperse system and applied radiation on the kinetics.

## 2. The Model

### 2.1. Quantum size effects in photoelectric emission

In this Section, we deal with conditions for photoelectron emission from nanoparticles. A quantum yield ($Y$) describing the ratio of the number of emitted photoelectrons to the number of incident photons (onto unit area) can be expressed as a function of the energy of incident photons of electromagnetic radiation by using the simplified form of the Fowler law given in [8]:

$$Y(\hbar\omega) = c(r_i)[\hbar\omega - W_e(r_i)]^2, \qquad (1)$$

where $c(r_i)$, and $W_e(r_i)$ are the photoemission constant and the work function, respectively, both dependent on the particle size (here the work function and the photon energy are expressed in eV).

The dependence of the work function on the size of the $i$-th particle includes the following terms [6,11]

$$W_e(r_i) = W_e - \Delta W_e(r_i) - \Delta W_F(r_i) + |e|(Z_i + |e|)/(4\pi\varepsilon_0\varepsilon r_i), \qquad (2)$$

where $W_e$ is the work function of the bulk, the terms $\Delta W_e(r_i) = (5/8) \cdot e^2/(4\pi\varepsilon_0 \varepsilon r_i)$ and $|e|(Z_i + |e|)/(4\pi\varepsilon_0 \varepsilon r_i)$ are size dependent additions to the work function of a charged spherical particle [11], $\varepsilon$ is the dielectric permeability of the interparticle medium ($\varepsilon = 1$ for gas media), $\varepsilon_0$ is the permittivity of free space, $e$ is the electron charge. For particles' shapes approximated by ideal spheres [6,9], the size-dependent addition to the Fermi energy is equal to

$$\Delta W_F(r_i) = W_F(r_i) - W_\infty \approx \pi W_\infty S_i / (4 k_F V_i). \tag{3}$$

where $k_F$ is the wave number at the Fermi level, $W_\infty = \hbar^2 (3\pi^2 n_e)^{2/3} (2m_e)^{-1} = (\hbar k_F)^2 (2m_e)^{-1}$ is the Fermi energy of the metal bulk ($n_e$ is the electron concentration, $m_e$ is the electron mass); $r_i$, $S_i$, $V_i$, are the radius, surface and volume of a particle, respectively. The importance of the size-dependent addition to the Fermi energy was emphasized in the paper [9].

The photoemission constant $c(r_i)$ for silver has been measured experimentally in [8] in the UV spectral range (in the range of photon energies near threshold). A strong size dependence of $c(r_i)$ in the range $2r_i = 4 - 6$ nm and an increase in the constant by a factor of 100 (compared to the bulk) was reported in [8] (see Figure 1).

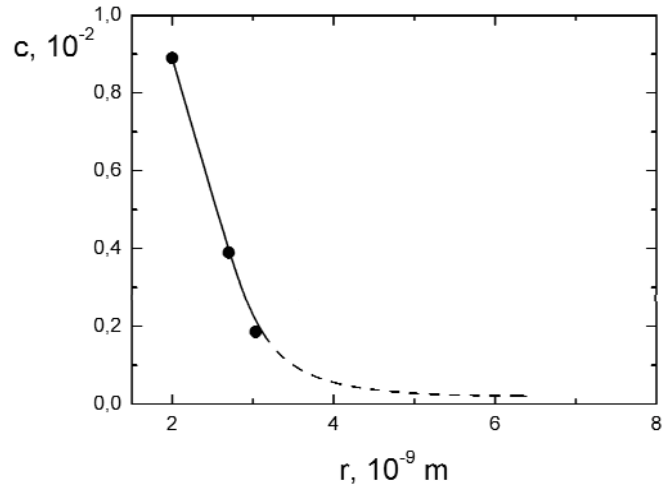

Figure 1. The dependence of the photoemission constant on the particle radius (in the range 2–3 nm) for silver at 230 nm (5.4 eV) (Schmidt-Ott et al 1980) (for larger radii the curve is a result of approximation).

The conventional values for the work function of *Ag* bulk (4.25–4.3 eV) were borrowed from [12,13]. It should be emphasized that for metals, the Fowler law in its form (1) is valid near the photoemission threshold of the photon energy only within the range $(\hbar\omega - W_e(r_i)) \leq (1.5 \div 2)$ eV. Beyond this range, dependence (1) can deviate from the square-law $(\hbar\omega - W_e(r_i))^2 \Rightarrow (\hbar\omega - W_e(r_i))^m$; the index $m \neq 2$.

In [14-16], the quantum yield $Y_w(\hbar\omega)$ dependence (in arbitrary units) was studied experimentally for nanoparticles in a wider spectral range (up to 10–11 eV) (see [15,16] for a study on *Ag* nanoparticles with a 5 nm radius). For *Ag* nanoparticles, this function $Y_w(\hbar\omega)$ as a correction to the Fowler law expression (1) can be taken into account by introducing a spectral-dependent term based on experimental data

$$F(\hbar\omega) = Y_w^0(\hbar\omega) / Y^0(\hbar\omega).$$

Here $Y_w^0(\hbar\omega)$ is the experimentally determined quantum yield (in relative units) in a wide range of photon energies covering the range of non-Fowler behavior

$Y^0(\hbar\omega)$ is the spectral dependence of the quantum yield (in relative units) as found from expression (1). In terms of the above definition, $F(\hbar\omega_{min}) \sim 1$ in the spectral range of validity of the Fowler law and deviates from 1 at higher photon energies.

We have thus harmonized the experimental and calculated quantum yield data obtained in different papers by bringing them to a common system of relative units in a wide spectral range. Taking into consideration these results, square-law expression (1), which is valid in a narrow spectral range, can be rewritten with the same index $m=2$ as follows:

$$Y_c(\hbar\omega, r_i) = F(\hbar\omega)c(r_i)[\hbar\omega - W_e(r_i)]^2. \qquad (4)$$

Expression (4) in this form is valid in the range up to 10 eV.

Within the range of the particle radii 2–5 nm, we assume the spectral term $F(\hbar\omega)$ (in relative units) to be independent of the particle size because the $Y(\hbar\omega)$ curves for the bulk and for a nanoparticle [14,15] are of the same shape. Expressions (1) and (4) similarly describe photoelectron emission at a small detuning of the photon energy from the red threshold, but at $(\hbar\omega - W_e(r_i)) \geq 3 - 3.5$ eV the effect may be significant, depending on both the type of metal and the surface contamination.

An example of a comparative diagram of the experimental and calculated absolute yields is shown in Figure 2 for the particle radius 5 nm.

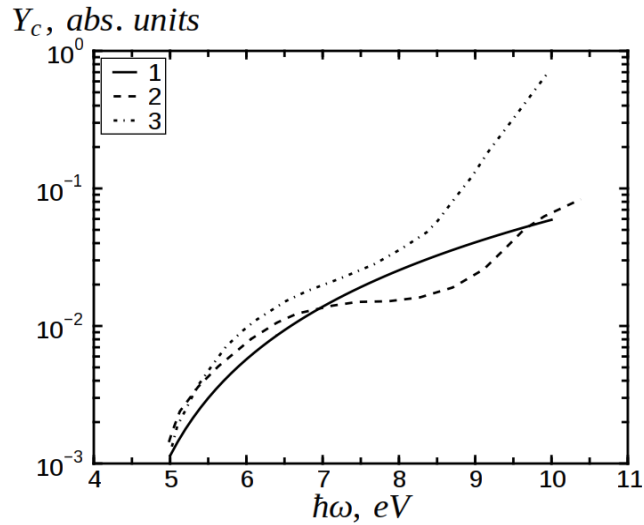

Figure 2. Spectral dependences of the absolute quantum yield for a silver nanoparticle with the radius $r_i=5$ nm: 1 – derived from expression (1); 2 – experimental dependence for clean particles, 3 – experimental dependence for contaminated nanoparticles (see comments in the text). The dependence $Y_c(\hbar\omega)$ calculated with expression (4) taking into consideration the experimental data reproduces the curve (3).

Curve 3 for contaminated particles abundant in the experiment (see [7,8,14-16]) was found to be the most realistic for further computations in our model. Similar curves can be calculated for other sizes of particles used in our study by taking into account the size dependent factors in Expression (4).

## 2.2. Mutual charging of different-size particles

### 2.2.1. Photoelectron emission

To investigate the aggregation process, we apply the method of molecular dynamics. Initially, all particles are randomly spaced in a rectangular cell with the sides about $10^2$–$10^3$ times the particle size. The walls of the cell can be assumed to be reflecting; that is considered as the boundary conditions. To simulate an infinite medium, we can change the boundary conditions to allow a particle irreversibly escape the cell while an identical particle with antiparallel momentum is simultaneously injected into the cell from the opposite side. As our computations showed, the use of both models provides similar results. Because of a simpler realization of the first model all computations in our work have been performed with the use of the reflecting wall cell.

Initial velocities are assigned to particles according to the Maxwell distribution. Next, mutual potential forces between the particles are introduced and discrete-time simulation of the Newtonian dynamics is performed (Section 2.3).

Emission of a given monovelocity $(V_e)_k$ fraction of photoelectrons ceases when the charge of the $j$-th particle reaches $Z_j$ and the photon energy is not sufficient to exceed the additional electrostatic energy of the particle after the emission of electrons with the total charge $Z_j$.

Velocities of the emitted photoelectrons should meet the following requirement $(V_e(\hbar\omega))_k = \sqrt{2[\hbar\omega - W_e(r_j)]/m_e}$ for the $k$-th electron and monochromatic irradiation. Estimates show that electron velocities at typical electron concentrations never become maxwellian, - neither during the time-step, nor at the initial stage of aggregation. In such situations, the probabilistic method is the most suitable tool for description of emission and capture of photoelectrons by particles.

To describe emission of the $k$-th electron by the $j$-th particle during the time-step $\Delta t$, a random process was realized. The requirement for emission from the $j$-th particle is $f \leq (P_{em})_j$. where $0 < f < 1$ is a random number chosen from the continuous uniform distribution, $(P_{em})_j$ is the probability of emission of an electron for a time step $\Delta t$. If $f \leq (P_{em})_j$, emission occurs. The probability is found as

$$(P_{em})_j = I_0 \cdot Y_c \cdot (S_p)_j \cdot \Delta t. \tag{5}$$

Restrictions on the parameters in this formula (primarily on the time-step) are set by the condition: $0 \leq (P_{em})_j \leq 1$. $I_0$ is the number of photons incident onto unit area of the particle surface $(S_p)_j$ during the time step. The emission dependence on the particle charge is included in the term $Y_c(W(r_j))$. When an electron is emitted from a particle, one elementary negative charge ($e$) is subtracted from the current value the charge of $j$-th particle ($Z_j$).

As shown in our paper [17] on the study of kinetics of aggregation during *half-coagulation time* ($t_{1/2}$), single-particle and two-particle sub-aggregates prevail while the share of three-particle aggregates is negligible (according to the definition (see e.g. [18]) $t_{1/2}$ is the time of coagulation within which the number of particles in the ensemble including sub-aggregates becomes half the initial number).

Thus, consideration of the emission and electron capture by particles is complicated by the presence of a relatively large number of two-particle sub-aggregates and by ohmic contacts of particles in aggregates. In this case, we are to solve a separate complex problem of emission and capture of electrons by a pair of particles with the charge ($Z_i+Z_j$) and non-isotropic electric potential. It is obvious, that we have to apply a simplified approach. Estimates show that in the case of a pair of particles the potential of a two-particle sub-aggregate with the particle radii $r_i$ and $r_j$ and charges $(Z_i+Z_j)$ is close to the potential of an equivalent spherical particle with the radius $r_c^{(i)} = \sqrt{r_i^2 + r_j^2}$ and the same charge $(Z_i+Z_j)$. After coagulation of *i*-th and *j*-th heteropolar particles the total charge $(Z_i+Z_j)$ is close to minimum. According to [19], the difference in the potential of two monopolarly charged spheres and the potential of such an equivalent sphere with radius $r_c^{(i)}$ and the same charge lies within 15%. This justifies our using a simplified method of calculation of emission and capture of electrons by two-particle sub-aggregates by way of substituting them with an equivalent spherical particle with the radius $r_c^{(i)}$ and charge $(Z_i+Z_j)$. This condition has been realized in our model.

Consideration of later stages of aggregation with a precise analysis of the potential of more complex multi-particle sub-aggregates in the calculation of emission and electron capture and, consequently, the effect of mutual charging on the aggregation is a separate problem and it is beyond the scope of our work.

### 2.2.2. Electron capture

Emitted electrons are randomly distributed in the interparticle space of a cell and are captured by particles due to collisions. Capture of photoelectrons depends on the sign of the particle charge. The maximum number of photoelectrons in the interparticle medium in a monovelocity fraction, which can be adsorbed by a particle, is limited because particles with negative charges repulse electrons.

Capture of the *k*-th electron by the *j*-th particle during the time-step $\Delta t$ is also realized as a random process by means of a random number ($0 < f < 1$) chosen from the continuous uniform distribution $\rho(x)$. In this case the probability (expectancy) of hitting of random number *f* in the interval $(x_{j-1}, x_j)$ is

$$P_j^{(k)} = P^{(k)}\left(x_{j-1}^{(k)} \leq f \leq x_j^{(k)}\right) = \int_{x_{j-1}^{(k)}}^{x_j^{(k)}} \rho(x)dx = x_j^{(k)} - x_{j-1}^{(k)}.$$

In other words, the condition for capture of the *k*-th electron by the *j*-th particle during time-step $\Delta t$ is

$$\sum_{l=1}^{l=j} P_l^{(k)} \geq f \geq \sum_{l=1}^{l=j-1} P_l^{(k)}, \tag{6}$$

where $P_j^{(k)}$ is the probability of electron capture by the *j*-th particle per time step $\Delta t$. If condition (6) is satisfied, the act of capture occurs.

To check this condition, the probability of capture of the *k*-th electron by the *j*-th particle for the time step $\Delta t$ is calculated from the expression

$$P_j^{(k)} = [\sigma_j^{(k)}((V_e)_k, \varphi_j) \cdot (V_e)_k / \upsilon_c]\Delta t. \tag{7}$$

$\upsilon_c$ is the cell volume.

The cross-section of capture of electrons [20] by the *j*-th particle is

$$\sigma_j^{(k)} = \begin{cases} \pi r_j^2 \left(1 + \dfrac{2e\varphi_j}{m_e(V_e)_k^2}\right), & \dfrac{2e\varphi_j}{m_e(V_e)_k^2} > -1 \\ 0, & \dfrac{2e\varphi_j}{m_e(V_e)_k^2} < -1 \end{cases},$$

where $\varphi_j$ is the surface potential of the *j*-th particle (values of $\varphi_j$ determine the limit of capture of electrons by a charged particle). Restrictions on the parameters in Formula (7) are also imposed by the condition $0 \leq P_j^{(k)} \leq 1$.

If an electron is captured, one elementary negative charge (*e*) is added to the current charge of the *j*-th particle ($Z_j$). The exchange of electrons emitted by different size particles continues until the charges of the particles prevent the emission and capture of electrons.

### *2.3. Kinetics of sol aggregation under mutual heteropolar charging*

Basic equations of the molecular dynamics method used for studying aggregation kinetics under interparticle electrostatic interactions are described in [2,17]. The resultant force acting on the *i*-th particle is subject to electrostatic influence of the remaining particles. This influence randomizes the motion of particles, giving rise to the appearance electrostatically linked particles with fast orbital rotation around the center of mass, which additionally intensifies the collision of particles. Coming into contact results in an irreversible coagulation of the particles. Irreversible coagulation of particles occurs when they come close to each other (approximately defined as $0.1(r_i + r_j)/2$) within the range of action of van-der-Waals forces. This provides for the sticking probability of particles in our model to be equal to 1. Brownian dynamics models taking into consideration stochastic forces and short-range repulsion of particles with adsorption layers are described in [21], where we analyze coagulation of particles in viscous media in hydrosols with the sticking probability of particles less than 1. However the use of this model for aerosols is inappropriate because in aerosols the Langevin equations for ensembles of particles are transformed into simple Newtonian equations (8).

In the absence of collisions, the coordinates and velocity of the *i*-th particle change during one iteration step according to the equations:

$$\frac{d\mathbf{r_i}}{dt} = \mathbf{v_i}, \quad m_i \frac{d\mathbf{v_i}}{dt} = \mathbf{F_i}, \quad \mathbf{F}_i = -\mathbf{grad}(U_{tot})_i, \tag{8}$$

where $\mathbf{r}_i$, $\mathbf{v}_i$, $m_i$, and $\mathbf{F}_i$ are the *i*–th particle coordinates, velocity, mass and the resultant force, respectively. The one-step Runge-Kutta method is used to solve the equations of motion of the particles. The force is calculated at each time moment *t* from the total interparticle interaction potential for all the pairs of particles which include an *i*-th particle (for the given size distribution function of particles). The pair-interaction energy $U_{tot}$ involves van-der-Waals [7,22] (9) and electrostatic (10) interactions:

$$U_v(r_{ij}) = -\frac{A_H}{6}\left(\frac{2r_i r_j}{r_{ij}^2 - (r_i + r_j)^2} + \frac{2r_i r_j}{r_{ij}^2 - (r_i - r_j)^2} + \ln\frac{r_{ij}^2 - (r_i + r_j)^2}{r_{ij}^2 - (r_i - r_j)^2}\right), \tag{9}$$

where, $A_H$ is the Hamaker constant ($A_H \approx 2\cdot 10^{-19}$ J for silver); $r_i$, and $r_j$, are the radii of particles, and $r_{ij}$ is the interparticle distance;

$$U_{el}(r_{ij}) = \frac{Z_i Z_j}{4\pi\varepsilon_0 \varepsilon r_{ij}}; \tag{10}$$

$Z_i$, $Z_j$ are the charges of interacting particles.

Distances between all pairs of particles are calculated during each iteration step. When the distance $r_{ij}$ between two particles *i* and *j* becomes equal to or less than the sphere diameter (when particles collide), these two particles are considered to be rigidly attached to each other, and the sub-aggregates so formed continue moving as a rigid body for the rest of the aggregation process under the influence of forces acting on each individual particle. Note that the particle sticking condition is checked after an elementary move. This means that technically they can overlap (the distance between the particle centers may become smaller than the sum of radii), but the time step is chosen to be sufficiently small (and velocity dependent) so that the depth of such overlapping is small compared to the sum of radii, and when overlapping happens, the interparticle distance is slightly increased to the point of exact touching.

Elementary rotation of a sub-aggregate (with more than one particle) is taken into account with the use of the rotation matrices

$$R_x^\alpha = \begin{pmatrix} 1 & 0 & 0 \\ 0 & \cos\alpha & -\sin\alpha \\ 0 & \sin\alpha & \cos\alpha \end{pmatrix}; \quad R_y^\beta = \begin{pmatrix} \cos\beta & 0 & \sin\beta \\ 0 & 1 & 0 \\ -\sin\beta & 0 & \cos\beta \end{pmatrix}; \quad R_z^\gamma = \begin{pmatrix} \cos\gamma & -\sin\gamma & 0 \\ \sin\gamma & \cos\gamma & 0 \\ 0 & 0 & 1 \end{pmatrix}. \quad (11)$$

These transformations are applied step by step to all particles in a given sub-aggregate. New coordinates resultant from the rotation are calculated as $\mathbf{r}'_i = R_x^\alpha R_y^\beta R_z^\gamma \mathbf{r}_i$, where $\alpha=\omega_x dt$, $\beta=\omega_y dt$, $\gamma=\omega_z dt$, and $\boldsymbol{\omega}$ is the angular velocity of an aggregate, which satisfies $\hat{I}\boldsymbol{\omega}=\mathbf{J}$, where $\mathbf{J}$ is the angular moment and $\hat{I}$ is the tensor of inertia of the aggregate being rotated. The angular velocity is calculated from the moment of inertia by inverting the tensor $\hat{I}$, namely, $\boldsymbol{\omega} = \hat{I}^{-1}\mathbf{J}$. Linear velocities of individual particles in an aggregate change as the result of rotation according to $\mathbf{v}'_i = \mathbf{v}_c + \boldsymbol{\omega}\times(\mathbf{r}_i - \mathbf{r}_c)$ where $\mathbf{r}_c$ is the center of mass of the aggregate. In turn, the angular momentum is updated according to $\mathbf{J}'=\mathbf{J}+\mathbf{M}dt$ where $\mathbf{M}$ is the total torque acting on the aggregate being rotated. At the starting instance, $\mathbf{M}=0$. For each subsequent iteration step, $\mathbf{M}$ is found from the known forces that act on each individual particle in the sub-aggregate: $\mathbf{M} = \mathbf{M} + d\mathbf{M}; d\mathbf{M} = \Sigma_i \mathbf{F}_i \times (\mathbf{r}_i - \mathbf{r}_c)$.

As mentioned above, particles are distributed in a cubic cell with a given volume concentration. Simulating particle coagulation in real time (aggregation time is on the order of tens of seconds and longer) with the time step $\Delta t = 10^{-8}$ s and preserving all parameters of the system would require a significant amount of computation time.

To speed up computations, an accelerated sol aggregation under mutual charging was imitated. The real aerosols studied in the experiment [7] had very low particle concentrations (on the order of $10^{10} - 10^{12} m^{-3}$) and hence the rate of spontaneous aggregation was very slow. In order to achieve acceleration within shorter times (less than $10^{-4}$ s), the concentration of the dispersed phase in the model has been increased to $10^{15}$ cm$^{-3}$ (see validation of this method in Section 3.2).

Another problem of computation is posed by the photoelectron flux from the particles being rather small at low light intensities close to the experimental value (less than 1 W/cm$^2$). Mutual charging of nanoparticles in a sol with a given concentration ceases when the rate of emission of photoelectrons and the inverse time of redistribution of photoelectrons among particles are below or comparable with the inverse half-time ($t_{1/2}^{-1}$) of particle coagulation.

These conditions impose the lower limit on the intensity of radiation that can initiate aggregation. Appearance of oppositely charged particles even in low-concentration aerosols essentially accelerates their aggregation in comparison with a spontaneous process. In computations, in order to compensate for insufficient photoelectron emission, the radiation intensity (or the density of the photon flux $I$) is increased by 3-4 orders of magnitude and more, which allows significant mutual charging and accelerated photoaggregation to be achieved within shorter computation times.

To speed up computations further, we used a time dependent dynamic step. The time step depends on the position of particles and is $10^{-9}$ maximum, averaging at $10^{-11}$ s. When particles are far from each other, the time step is maximal, as the particles approach each other (even if in just one pair) the step shortens.

## 3. Results and Discussion

In this Section we present the results of computation of the coagulation kinetics employing the model described in Section 2 and analyze the influence of different characteristics of the disperse system and radiation on the kinetics.

Maxwellian velocities of particles in random directions are found depending on the aerosol temperature. As mentioned above, a polydisperse ensemble comes to equilibrium because photoelectron emission depends on the particle size. Electrons are randomly redistributed among particles causing mutual heteropolar charging of the particles. As mentioned in Section 2.3, to verify this effect, we have calculated electric charge distribution histograms for 100 silver particles ranging in size from 4 to 6 nm with the Poisson distribution (see figure 3).

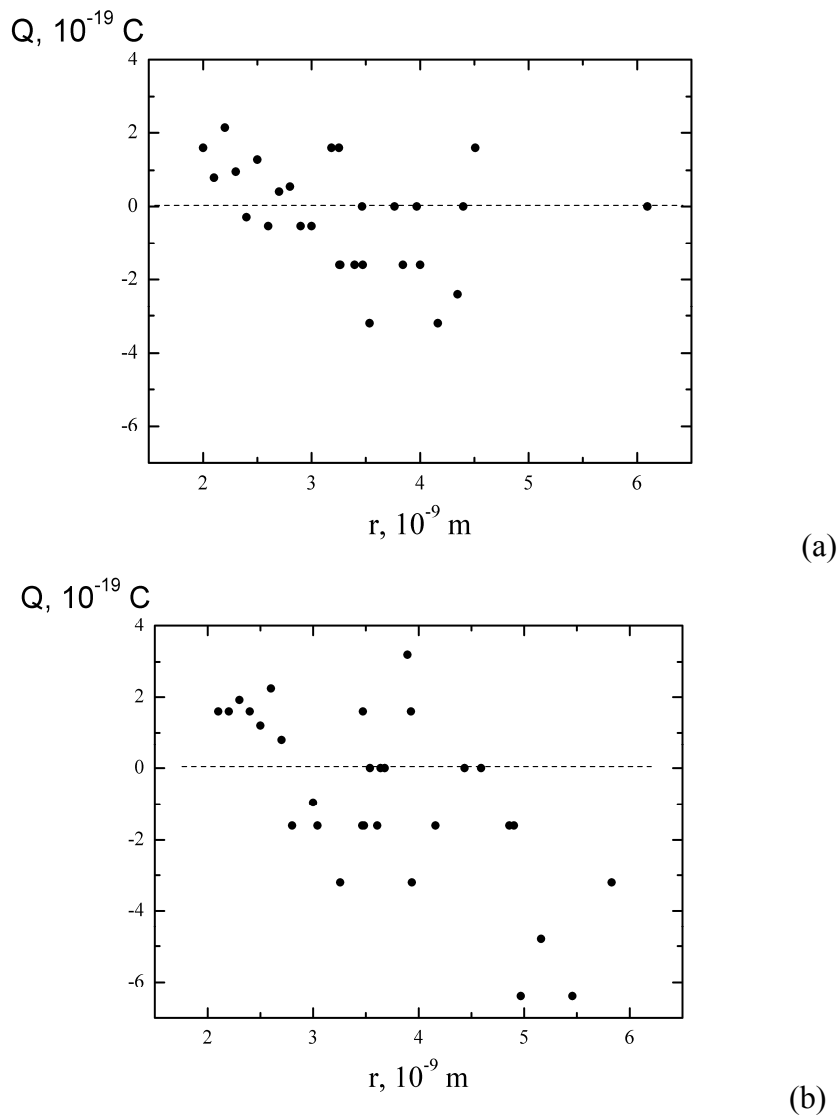

Figure 3. The histogram of distribution of an electric charge on 100 *Ag* particles of a polydisperse ensemble for the Poisson distribution of particle sizes after UV irradiation. (a) corresponds to the energy of photons 6 eV and (b) corresponds to the photon energy 8 eV. The radiation intensity is $I=10^6$ Wm$^{-2}$; the particle concentration is $n=10^{21}$ m$^{-3}$; half-coagulation time $t_{1/2}$ is approximately equal to $10^{-5}$ s. The ratio of the maximal and minimal radii of particles for the Poisson size distribution function is $r_{min}/r_{max}=2/3$, $r_{min}=2$ nm.

When the above described conditions are realized, we observe a significant exchange of electrons in a polydisperse ensemble of metal nanoparticles. Histograms of charges in Ag nanoparticles of a polydisperse ensemble are shown in Figure 3 for half-coagulation time. The emission of electrons continues as electrons are redistributed among the particles.

Calculations of charge distribution histograms for different size particles (the value of the charge in particles of a given size) under mutual charging of the particles are made for both single particles and multi-particle sub-aggregates treated as equivalent spheres. As mentioned above, upon coagulation of heteropolarly charged particles their charges undergo redistribution due to the ohmic contact that can be interpreted as the appearance of equivalent particles with the radius $r_c^{(i)} = \sqrt{r_i^2 + r_j^2}$ and total charge ($Z_i+Z_j$).

### *3.1. Coagulation kinetics*

The particles were irradiated by UV light with the photon energy 6 eV (Figure 3a). The same histogram for the photon energy 8 eV is shown in Figure 3b. The range larger than 6 nm corresponds to multi-particle sub-aggregates. These figures show that the mutual charging effect really takes place. A strong dispersion of the charge values is explained by a random nature of the process.

The photon energies for our simulations have been chosen arbitrary but with some justifications. The energy 6 eV corresponds to the spectral range of validity of the Fowler law $\hbar\omega - W_e(r_i) = 1.7$ eV and it is close to the photon energy in the experiments by [7,8]. The value 8 eV is outside this range but within the range of transparency of solids and yet the experimental realization does not require any sophisticated means of registration of photostimulated aggregation of aerosols under such conditions.

The difference in 6 eV (a) and 8 eV (b) histograms lies in that in Case (b) there are particles with larger charges and a larger dispersion of particle charges is observed because following expression (4) the higher the energy of photons, the larger the particle charge. But even under such conditions we register smaller particles to be charged positively and the larger ones to be charged negatively.

Predominance of the fraction of negatively charged particles is due to small particles emitting more electrons (see the size-dependence of the photoemission constant in Figure 1) and larger ones capturing more electrons.

In the process of aggregation, emission and capture of electrons are realized both for single particles and particles in sub-aggregates. Changes in the histograms are observed for large particles due to the capture of electrons by multi-particle sub-aggregates. Some of them have a charge close to zero.

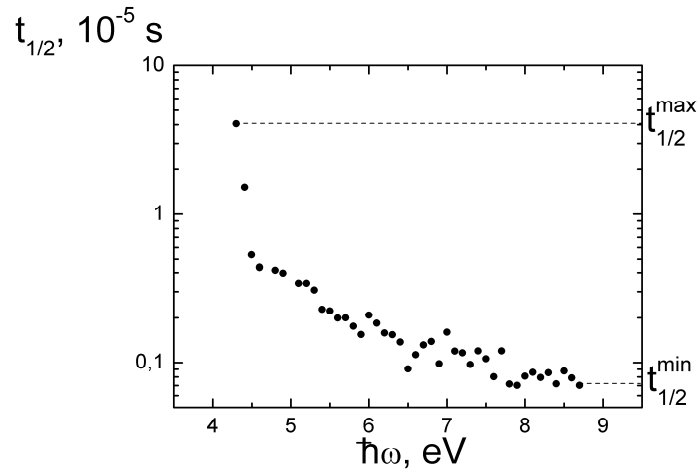

Figure 4. Dependence of the rate of photostimulated aggregation of a polydisperse ensemble of 100 silver nanoparticles on the energy of photons under monochromatic irradiation. $I=10^6$ Wm$^{-2}$, the particle concentration is $n=10^{21}$ m$^{-3}$, $r_{min}/r_{max}=2/3$, $r_{min}=2$ nm. $t_{1/2}^{max}=4.5 \cdot 10^{-5}$ s, $t_{1/2}^{min}=7.3 \cdot 10^{-7}$ s.

The rate of photostimulated aggregation of a polydisperse ensemble of 100 silver nanoparticles with the radius 2–3 nm is shown in Figure 4 as a function of the photon energy of monochromatic UV radiation.

Saturation at high photon energies results from the electrostatic limitation on the maximum charge of particles in acts of emission and capture of electrons, which in turn limits the interparticle Coulomb attractive forces.

Hereinafter the acceleration factor can be found from the ratio $t_{1/2}^{max}/t_{1/2}^{min}$ of maximal and minimal coagulation times given in the figure captions.

Dependence of the rate of photoaggregation of an ensemble of *Ag* nanoparticles on the the particle size distribution function is shown in Figure 5 for the bimodal and Poisson size distribution functions and 6 eV photon energy of monochromatic radiation complying with the condition $(r_{max}+r_{min})/2=3$.

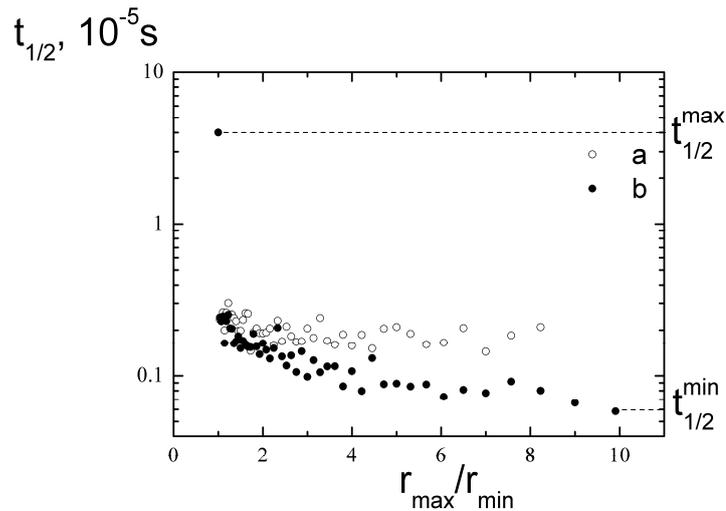

Figure 5. Dependence of the rate of photostimulated aggregation of a polydisperse ensemble of 100 silver nanoparticles on the particle size distribution function: (a) Poisson distribution and (b) bimodal distribution. The photon energy is 6 eV. ($I=10^6$ Wm$^{-2}$, $n=10^{21}$ m$^{-3}$; the average radius in a pair ($r_{max}, r_{min}$) is kept constant at 3 nm; $t_{1/2}^{max}=4.5\cdot 10^{-5}$ s, $t_{1/2}^{min}=6.7\cdot 10^{-7}$ s.

The dependence exhibits a tendency to saturation due to saturation of the dependence of the photoemission constant on the particle size (Figure 1): the larger the particle, the smaller the photoemission constant.

Both dependences in Figure 5 look similar: the coagulation time is the longest at equal particle sizes and rapidly shortens with the growing difference in the particle sizes when the mutual charging effect is maximal (points with $t_{1/2}^{max}$ for cases (a) and (b) coincide).

Figure 6 shows the rate of photostimulated aggregation of a polydisperse ensemble of silver nanoparticles irradiated by monochromatic light depending on the light intensity (the photon energy is 6 eV). This dependence demonstrates saturation at higher intensities at the given wavelength. Saturation at high intensities is explained by the electrostatic limitation on the maximal charge of particles.

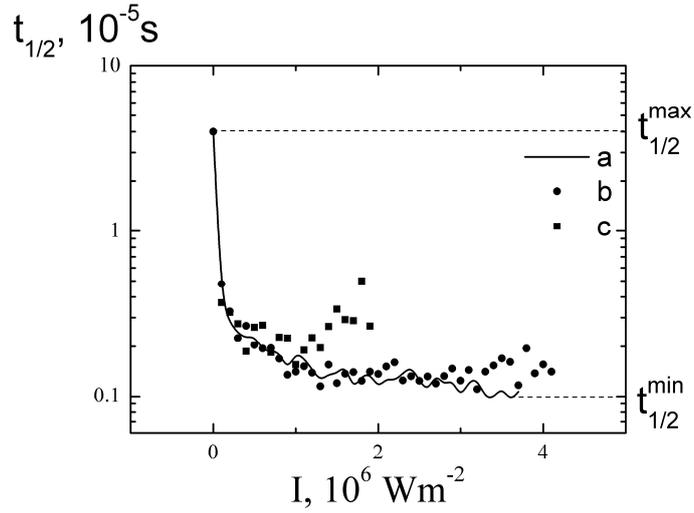

Figure 6. Dependence of the half-coagulation time of a polydisperse ensemble of silver nanoparticles irradiated by monochromatic light on the radiation intensity for different fractions of captured electrons (a – 1, b – 0.5, and c – 0.2). The photon energy is 6 eV, the particle concentration $n=10^{21}$ m$^{-3}$, the minimal radius of a particle is 2 nm, and the particle size distribution function is Poissonian ($r_{min}/r_{max}=2/3$, $r_{min}=2$ nm). $t_{1/2}^{max}=4.0\cdot 10^{-5}$ s, $t_{1/2}^{min}=9.5\cdot 10^{-7}$ s.

At too high radiation intensities, some electrons are not captured by particles and remain in the interparticle medium. Curve (a) corresponds to all electrons having been captured, (b) and (c) to 50% and 20% captured electrons, respectively. As we can see from these dependences, the lower the percentage of captured electrons, the slower the aggregation processes. This is due to a larger fraction of positively charged particles.

### *3.2. Limitations of the model*

Kinetics of photostimulated aggregation depending on the particle concentration is shown in Figure 7.

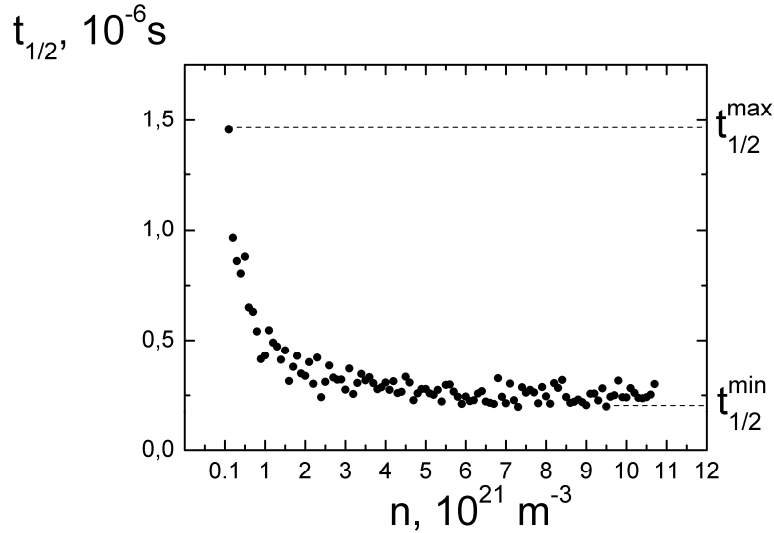

Figure 7. Dependence of the half-coagulation time of a polydisperse ensemble of silver nanoparticles on the particle concentration. The particles are irradiated by monochromatic light with the photon energy 6eV, $I=10^6$ Wm$^{-2}$. $r_{min}/r_{max}=2/3$, $r_{min}=2$ nm, $t_{1/2}^{max}=1.5\cdot 10^{-6}$ s, $t_{1/2}^{min}=2\cdot 10^{-7}$ s.

The revealed tendency agrees with the theoretical models derived from the Smoluchowski theory. In concordance within this theory, the dependence of the sol coagulation rate on the particle concentration ($t_{1/2} \sim n^{-1}$) (see, e.g., [18]) valid for arbitrary particle concentration in random systems is given by the equation

$$t_{1/2} = \frac{6\eta r}{4\rho n k_B T}. \tag{12}$$

In the case of a rarefied interparticle gas medium and nanoscopic particles an approximate estimate can be provided by the expression [18,23]

$$t_{1/2} \approx \frac{6\eta r}{4\rho n k_B T \left(1 + A\frac{\lambda_0}{r}\right)}, \tag{13}$$

where $\eta$ is the coefficient of viscosity of the interparticle medium, $r$ is the radius of particles, $n$ is the initial particle concentration, $\rho$ is the range of action of interparticle forces, $A$ is an experimentally determined constant of the order of 1, $\lambda_0$ is a mean free path of gas molecules. For the van-der-Waals forces responsible for spontaneous aggregation the ratio $(\rho/r)_v \approx 2.3$. The range of action of electrostatic attractive forces can be found from the equation

$$\frac{Z_i Z_j}{4\pi\varepsilon_0\varepsilon\rho} = (3/2)k_B T. \tag{14}$$

Here $Z_i$ and $Z_j$ are charges of particles. So for the particle charges $Z_i$ and $Z_j$ being $\pm 1e$, $\pm 2e$, and $\pm 3e$, we have the ratio $(\rho/r)_{el} \approx$ 38, 152 and 342. In this case, a rough estimate of the acceleration factor of aggregation in comparison with a spontaneous process for the same particle radii $r$, with the unit sticking probability of particles can be found as the ratio $[(\rho/r)_{el}]/[(\rho/r)_v]$ or $[(t_{1/2})_{el}]/[(t_{1/2})_v]$. For charges $\pm 1e$, $\pm 2e$, $\pm 3e$ it yields an acceleration factor of 16, 66, and 148 that is close to the computation results obtained by the molecular dynamics method.

According to expressions (12) and (13), the acceleration factor is independent of the particle concentration. This means our method can be applied to investigate systems with much lower particle concentrations typical of experimental conditions ($10^{10}$–$10^{12}$ m$^{-3}$) [7]. Note that in vacuum interparticle media, randomization of charged particles is caused by their electrostatic interactions.

The above results have been obtained for the case when the mean free path of emitted photoelectrons ($l_e$) exceeds both the mean interparticle distances ($l_p$) and the cell size. In this case, photoelectrons fill the cell as a continuous homogenous background. Obviously, if an electron reaches a neighboring particle within the time between collisions of particles, the effect of mutual charging may occur. If the mean free path decreases, the effect is less manifest.

In real sols, a dense interparticle medium reduces the mean free path of photoelectrons because of scattering on molecules. For example, in water the mean free path of photoelectrons with energy 4 eV does not exceed 5 nm [24].

In simulating this process, one of the simplest ways to describe the slowdown of electron propagation in the interparticle medium is to introduce the factor ($V_m$) into the real photoelectron velocity $V_e$ and to substitute $V_e$ by $V_m \times V_e$ for emitted photoelectrons.

Figure 8 demonstrates the dependence of the half-coagulation time on the slowing factor $V_m$ ($0 < V_m < 1$).

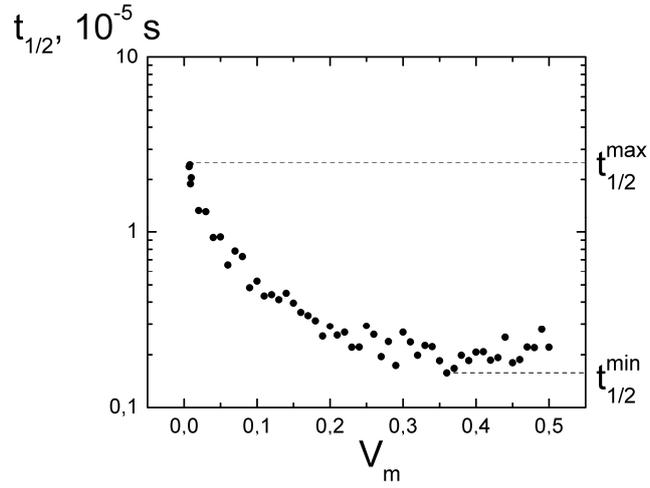

Figure 8. Dependence of the half-coagulation time on the slowing factor of photoelectrons. The particle concentration is $10^{21}$ m$^3$, $t_{½} = 4.6 \cdot 10^{-5}$ s for the Poisson size distribution function, $r_{min}/r_{max}=2/3$, $r_{min=}2$ nm. $I=10^6$ Wm$^{-2}$ and the photon energy is 6 eV. $t_{1/2}^{max}=2.5\cdot 10^{-5}$ s, $t_{1/2}^{min}=1.6 \cdot 10^{-6}$ s.

It can be seen that the half-coagulation time increases with decreasing $V_m$. This is due to recapture of the emitted electron by the same particle. The cross-section of the electron capture by the $j$-th particle is $\sigma_j^{(i)} = \pi r_j^2 \left(1+\frac{2e\varphi_j}{m_e(V_e)_i^2}\right)$. Following this expression, $\sigma_j^{(i)}$ increases as $V_e$ reduces (if $\varphi_j > 0$). Irradiated particles tend to minimize their charges. In the limit it corresponds to the rate of spontaneous aggregation.

At high radiation intensities some fraction of electrons is not captured by particles (Figure 9) and remains in the interparticle medium, which is another contributing factor to the slowing of the aggregation process.

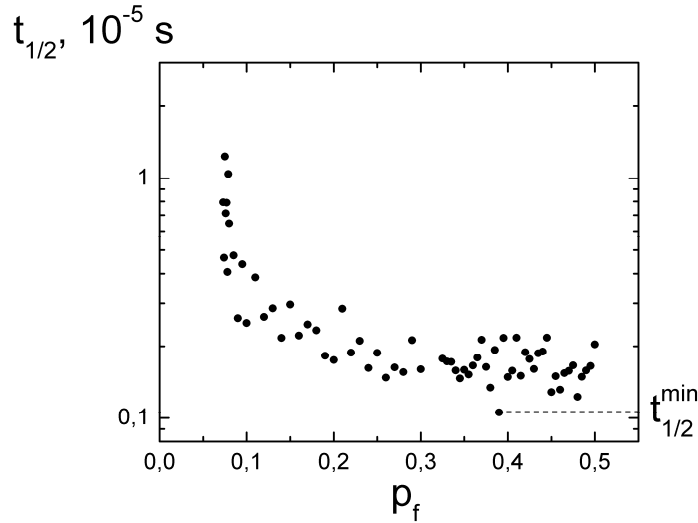

Figure 9. Dependence of the half-coagulation time on the fraction ($p_f$) of photoelectrons captured by particles. The ratio of radii of Ag particles $r_{min}/r_{max}=2/3$, the particle concentration is $10^{21}$ m$^3$, $t_{½} = 4.6\cdot 10^{-5}$ s for the Poisson size distribution function, $r_{min}/r_{max}=2/3$, $r_{min=}2$ nm., $I=10^6$ Wm$^{-2}$ and photon energy 6 eV. $t_{1/2}^{max} \to \infty$, $t_{1/2}^{min}=10^{-6}$ s.

Computations show that as the number of captured electrons reduces, the half-coagulation time indefinitely grows because of domination of particles with the positive charge.

## Conclusion

We can conclude that the proposed model can be applied for studying light accelerated coagulation of real disperse systems with various concentrations of particles as well as rarefied gaseous or vacuum interparticle media. We hope this model could provide a useful tool to predict the effect of heteropolar mutual charging of nanoparticles due to the size dependent photoelectric effect in polydisperse metal sols and what is even more important in heterosols with an interparticle medium transparent for photoelectrons. It has been shown that under considered conditions the mutual charging effect can accelerate photoaggregation of Ag aerosols up to 60 times and more in comparison with the rate of their spontaneous aggregation without light.

An accelerated aggregation of silver aerosols with nanoparticles of various sizes was observed in the experiments [7] in a $N_2$ flow. The nanoparticles were produced by electrode sputtering in a spark discharge between *Ag* electrodes accompanied by generation of plasma. This experimentally established fact of an anomalously rapid aggregation of metal aerosols remains unexplained so far. The paper [9] has been the first to draw attention to a possibility of explaining the results of [7] by mutual charging of the particles. In our opinion, one of possible explanations for the accelerated aggregation of aerosols in that paper should be sought in the experimental conditions. In that experiment, the concentrations of particles with different sizes were determined by the method of photoelectron emission from the obtained nanoparticles under UV irradiation. Note also that the aerosol used in the experiment was not strictly monodisperse. The contribution of these two factors to the aggregation kinetics can provide mutual charging of the particles.

Optimal conditions for manifestation of the mutual charging effect primarily involve enrichment of the radiation spectrum by high-energy photons (over 10 eV), which eases restrictions on the values of particle charges, enhances the Coulomb interparticle attraction forces and further accelerate aggregation. This concerns observation of the effect in aerosols under natural conditions, under which the photoemission characteristics of particles may be associated not only with the size dependence, but also with the chemical composition of particles. At high intensities of pulsed laser fields, an accelerated aggregation of aerosols may be induced by multi-photon electron photoemission. Another important factor is the presence of particles in metal aerosols with a diameter smaller than 10 nm because the photoemission constant of such particles increases rapidly with the decreasing particle size.

Our paper demonstrates how fine quantum size effects in metal nanoparticles can influence the aggregation kinetics, and we would like to draw attention to the possibility of manifestation of this effect.

**Acknowledgments**

This work was supported by grants from the following foundations of the Russian Federation: State contract 02.740.11.0220, Presidium of RAS № 27.1, OFN RAS № 9.1, SB RAS № 5, NS-7810.2010.3, RNP VS 2.1.1.1814.